\documentclass[nofootinbib,prd,showpacs]{revtex4}
\usepackage{amsmath}
\usepackage{subfigure}
\usepackage{amsfonts}
\usepackage{amssymb}
\usepackage{graphicx}
\graphicspath{ {figs/} }

\usepackage[latin1]{inputenc}
\usepackage[english]{babel}
\usepackage{color,xcolor}

\definecolor{purple}{rgb}{1,0,1}
\definecolor{lime}{HTML}{A6CE39} 

\definecolor{lime}{HTML}{A6CE39}

\begin{document}
\title{
Comment on ``Source of black bounces in Rastall gravity''}


\author{Manuel E. Rodrigues}
\email{esialg@gmail.com}
\affiliation{Faculdade de Ci\^{e}ncias Exatas e Tecnologia, 
Universidade Federal do Par\'{a}\\
Campus Universit\'{a}rio de Abaetetuba, 68440-000, Abaetetuba, Par\'{a}, 
Brazil}
\affiliation{Faculdade de F\'{\i}sica, Programa de P\'{o}s-Gradua\c{c}\~ao em 
F\'isica, Universidade Federal do 
 Par\'{a}, 66075-110, Bel\'{e}m, Par\'{a}, Brazil}

 \author{Marcos V. de S. Silva}
\email{marco2s303@gmail.com}
\affiliation{Departamento de F\'isica, Programa de P\'os-Gradua\c c\~ao em F\'isica, Universidade Federal do Cear\'a, Campus Pici, 60440-900, Fortaleza, Cear\'a, Brazil}
\affiliation{Faculdade de F\'{\i}sica, Universidade Federal do Par\'{a},
 68721-000, Salin\'opolis, Par\'{a}, Brazil}
\date{today; \LaTeX-ed \today}
\begin{abstract}
In Atazadeh and Hadi (JCAP \textbf{01}, 067 (2024)), the authors proposed that black bounce solutions, such as the Simpson-Visser and the Bardeen-type spacetimes, can be obtained from Rastall gravity. To achieve these spacetimes, the authors consider the presence of a phantom scalar field with nonlinear electrodynamics. However, in this comment, we obtained different electromagnetic Lagrangians from the original work. The most problematic issue is not the incorrect expression of the electromagnetic Lagrangian itself. We show that the method obtains electromagnetic functions that are inconsistent. To obtain the black bounce spacetimes as solutions of Rastall gravity, it is necessary to consider an isotropic fluid, combined with the nonlinear electrodynamics and the phantom scalar field.
\end{abstract}
\maketitle
\def\HMS{{\scriptscriptstyle{HMS}}}
\section{Field Equations}
\label{S:FieldEq}
Following the same definitions of the original work \cite{Atazadeh:2023wdw}, the field equations are
\begin{eqnarray}
 &&   \nabla_\mu \left[L_F F^{\mu\nu}\right]=\frac{1}{\sqrt{-g}}\partial_\mu \left[\sqrt{-g}L_F F^{\mu\nu}\right]=0\label{electro},\\
 &&   2\epsilon \nabla_\mu \nabla^\mu\phi=-\frac{dV(\phi)}{d\phi},\label{scalar}\\
   && R_{\mu\nu}-\frac{1}{2}g_{\mu\nu}R+\gamma g_{\mu\nu}R=\kappa^2T^{\phi}_{\mu\nu}+\kappa^2T^{EM}_{\mu\nu},\label{einstein}
\end{eqnarray}
where $L_F=\partial L/\partial F$, with the Lagrangian $L(F)$ being a general function of the electromagnetic scalar $F=F^{\mu\nu}F_{\mu\nu}/4$, $\phi$ is a scalar field, $V(\phi)$ is the potential related to the scalar field, $R$ is the trace of the Ricci tensor $R_{\mu\nu}$, $g$ is the determinate of the metric $g_{\mu\nu}$, $\lambda$ is the Rastall parameter, $T^{\phi}_{\mu\nu}$ is the stress-energy tensor of the scalar field, and $T^{EM}_{\mu\nu}$ is the stress-energy tensor of the electromagnetic field. The stress-energy tensors are given by
\begin{eqnarray}
    T^{\phi}_{\mu\nu}&=&2\epsilon \partial_\mu\phi\partial_\nu\phi-g_{\mu\nu}\left(\epsilon g^{\alpha\beta}\partial_\alpha \phi \partial_\beta \phi-V(\phi)\right),\\
    T^{EM}_{\mu\nu}&=&g_{\mu\nu}L(F)-L_F {F_\mu}^\alpha F_{\nu \alpha}.
\end{eqnarray}

The form of the line element is written as
\begin{equation}
    ds^2=f(r)dt^2-f(r)^{-1}dr^2-\sigma^2 (r) \left(d\theta^2+\sin^2\theta d\varphi^2\right).\label{line}
\end{equation}

A magnetic field written as $F_{23}=Q \sin\theta$, where $Q$ is the magnetic charge, identically satisfies the equation \eqref{electro}.

The equation \eqref{einstein}, as done by the authors in \cite{Atazadeh:2023wdw}, can be rewrite in terms of the trace of the stress-energy tensor, $T={T^\mu}_\mu$. However, the form of the equation \eqref{einstein} will no affect the results.

The field equations, to the line element \eqref{line}, are written as,
\begin{eqnarray}
    -2 \epsilon  \left(f' \phi '+f \phi ''\right)-\frac{4 \epsilon  f \sigma ' \phi '}{\sigma }=-\frac{V'}{\phi '},\label{eqmov1}\\
   \gamma  f''+\frac{(4 \gamma -1) f' \sigma '}{\sigma }+\frac{2 (2 \gamma -1) f \sigma ''}{\sigma }+\frac{(2 \gamma -1) \left(f \sigma '^2-1\right)}{\sigma ^2}=\kappa ^2 \left(\epsilon  f \phi '^2+L+V\right),\label{eqmov2}\\
   \gamma  f''+\frac{4 \gamma  f' \sigma '}{\sigma }-\frac{f' \sigma '}{\sigma }+\frac{4 \gamma  f \sigma ''}{\sigma }+\frac{2 \gamma  f \sigma '^2}{\sigma ^2}-\frac{f \sigma '^2}{\sigma ^2}-\frac{2 \gamma }{\sigma ^2}+\frac{1}{\sigma ^2}=\kappa ^2 \left(L-\epsilon  f \phi '^2+ V\right),\label{eqmov3}\\
  \gamma  f''-\frac{f''}{2}+\frac{4 \gamma  f' \sigma '}{\sigma }-\frac{f' \sigma '}{\sigma }+\frac{4 \gamma  f \sigma ''}{\sigma }+\frac{2 \gamma  f \sigma '^2}{\sigma ^2}-\frac{f\sigma ''}{\sigma }-\frac{2 \gamma }{\sigma ^2}=\kappa ^2\left( L-\frac{ Q^2 L_F}{\sigma ^4}+ \epsilon  f \phi '^2+ V\right).\label{eqmov4}
\end{eqnarray}
Thus, we have a set of equations that can be used to find the sources of black hole bounce solutions, once the metric coefficients, $f(r)$ and $\sigma(r)$, are known.
\subsection{Simpson-Visser case}
The first case considered by the authors was the Simpson-Visser metric, whose metric coefficients are written as
\begin{equation}
    f(r)=1-\frac{2m}{\sqrt{r^2+Q^2}}, \qquad \sigma(r)=\sqrt{r^2+Q^2}.
\end{equation}
Substituting the metric coefficients in equations \eqref{scalar}, considering a phantom scalar field, $\epsilon=-1$ and integrating the necessary equations, we find
\begin{eqnarray}
  \phi=\frac{\tan ^{-1}\left(\frac{r}{Q}\right)}{\kappa }, \qquad V(r)=\frac{4 m Q^2}{5 \kappa ^2 \left(Q^2+r^2\right)^{5/2}},\label{Scalar_SV}\\
  L(r)=\frac{6 m Q^2}{5 \kappa ^2 \left(Q^2+r^2\right)^{5/2}}+\frac{2 \gamma  Q^2 \left( \sqrt{Q^2+r^2}-3 m\right)}{ \kappa ^2 \left(Q^2+r^2\right)^{5/2}}, \qquad L_F(r)=\frac{3 m}{\kappa ^2 \sqrt{Q^2+r^2}}.\label{Eletro_SV}
\end{eqnarray}
Comparing equation \eqref{Eletro_SV} with equation (33) in \cite{Atazadeh:2023wdw}, we verify that the results are not the same. Actually, there is a problem that arises when consider a nonlinear electrodynamics as a source. Despite we consider $L$ and $L_F$ as independent functions in the field equations, they are related by 
\begin{equation}
    L_F=\frac{\partial L}{\partial F}=\frac{\partial L}{\partial r}\left(\frac{\partial F}{\partial r}\right)^{-1} \longrightarrow L_F-\frac{\partial L}{\partial r}\left(\frac{\partial F}{\partial r}\right)^{-1}=0.\label{cons}
\end{equation}
However, booth results, \eqref{Eletro_SV} and (33) in \cite{Atazadeh:2023wdw}, do not satisfy this condition. Actually, if we integrate $L_F$ in order to obtain the Lagrangian $L$, we find that
\begin{equation}
    L(r)=\frac{6 m Q^2}{5 \kappa ^2 \left(Q^2+r^2\right)^{5/2}},
\end{equation}
which is the same result from \eqref{Eletro_SV} with out the correction from the Rastall gravity.

\subsection{Bardeen-type}
The second model consider in \cite{Atazadeh:2023wdw} is the Bardeen-type black bounce, given by
\begin{equation}
    f(r)=1-\frac{2mr^2}{\left(r^2+Q^2\right)^{3/2}}, \qquad \sigma(r)=\sqrt{r^2+Q^2}.
\end{equation}
Doing the same process that we did to the Simpson-Visser spacetime, we find
\begin{eqnarray}
    \phi(r)&=&\frac{\tan ^{-1}\left(\frac{r}{Q}\right)}{\kappa }, \qquad V(r)=\frac{4 m Q^2 \left(7 r^2-8 Q^2\right)}{35 \kappa ^2 \left(Q^2+r^2\right)^{7/2}},\\
    L(r)&=&\frac{2 m Q^2 \left(16 Q^2+91 r^2\right)}{35 \kappa ^2 \left(Q^2+r^2\right)^{7/2}}-\frac{2 \gamma  m Q^2 \left(2 Q^2+r^2\right)}{\kappa ^2 \left(Q^2+r^2\right)^{7/2}}+\frac{2 \gamma  Q^2}{\kappa ^2 \left(Q^2+r^2\right)^2}, \quad L_F(r)=\frac{m \left(13 r^2-2 Q^2\right)}{\kappa ^2 \left(Q^2+r^2\right)^{3/2}}.
\end{eqnarray}
Similarly to what happened in the previous case, the Lagrangian we obtained does not match the equation (40) from \cite{Atazadeh:2023wdw}. Furthermore, none of the Lagrangians satisfy the consistency relation \eqref{cons}. Integrating $L_F$, we find
\begin{equation}
    L(r)=\frac{2 m Q^2 \left(16 Q^2+91 r^2\right)}{35 \kappa ^2 \left(Q^2+r^2\right)^{7/2}},
\end{equation}
which does not consider the correction from the Rastall gravity.

\section{Anisotropic fluids}
Despite the problem in mapping the solution to a nonlinear electrodynamics with scalar field, we can still analyze it in the form of an anisotropic fluid.

The stress-energy tensor to an anisotropic fluid is given by \cite{Menchon:2017qed}
\begin{equation}
 {T^{\mu}}_{\nu} = (\rho + p_t)u^{\mu}u_{\nu} - p_t \delta^{\mu}_\nu + \left(p_r-p_t\right)\chi^\mu\chi_\nu,
\end{equation}
where $\rho$ is the energy density, $p_r$ is the radial pressure, and $p_t$ is the tangential pressure. The normalized vectors $u^\mu$ and $\chi^\mu$ are timelike, $u^\mu u_\mu=1$, and spacelike , $\chi^\mu \chi_\mu=-1$, respectively. 

In a comoving coordinate, $u^1=u^2=u^3=0$ and $\chi^0=\chi^2=\chi^3=0$, the stress-energy tensor can be express as \cite{Bayin:1985cd}
\begin{equation}
    {T^\mu}_\nu= \mbox{diag} \left[\rho,-p_r,-p_t,-p_t\right].\label{Aniso}
\end{equation}

Imposing the field equations to the Rastall gravity, \eqref{einstein}, and considering the stress-energy tensor to the anisotropic fluid, we find

\begin{eqnarray}
    &&\rho(r)=\frac{\gamma  \sigma ^2 f''(r)+\sigma  \left((4 \gamma -1) f' \sigma '+2 (2 \gamma -1) f \sigma ''\right)+(2 \gamma -1) \left(f \sigma
   '^2-1\right)}{\kappa ^2 \sigma ^2},\\
  && p_r(r)=\frac{2 \gamma -\gamma  \sigma ^2 f''+\sigma  \left((1-4 \gamma ) f' \sigma '-4 \gamma  f\sigma ''\right)+(1-2 \gamma ) f \sigma '^2-1}{\kappa
   ^2 \sigma ^2},\\
  && p_t(r)=\frac{4 \gamma +(1-2 \gamma ) \sigma ^2 f''-2 (4 \gamma -1) \sigma  \left(f' \sigma '+f \sigma ''\right)-4 \gamma  f \sigma '^2}{2 \kappa ^2
   \sigma ^2}.
\end{eqnarray}

If we want to find the fluid quantities that differ from the usual black bounces in general relativity, we calculate
\begin{eqnarray}
    &&\rho^R=\rho-\rho^{EM}-\rho^{\phi},\qquad p_r^R=p_r-p_r^{EM}-p_r^{\phi},\qquad p_t^R=p_t-p_t^{EM}-p_t^{\phi}.
\end{eqnarray}
\subsection{Simpson-Visser}
Considering again the Simpson-Visser spacetime, the components of the anisotropic fluid are written as
\begin{eqnarray}
    \rho=\frac{\gamma  Q^2 \left(2 \sqrt{Q^2+r^2}-6 m\right)}{\kappa ^2 \left(Q^2+r^2\right)^{5/2}}+\frac{Q^2 \left(4 m-\sqrt{Q^2+r^2}\right)}{\kappa ^2 \left(Q^2+r^2\right)^{5/2}},\\
    p_r=-\frac{\gamma  Q^2 \left(2 \sqrt{Q^2+r^2}-6 m\right)}{\kappa ^2 \left(Q^2+r^2\right)^{5/2}}-\frac{Q^2}{\kappa ^2 \left(Q^2+r^2\right)^2},\\
    p_t=\frac{\gamma  Q^2 \left(6 m-2 \sqrt{Q^2+r^2}\right)}{\kappa ^2 \left(Q^2+r^2\right)^{5/2}}+\frac{Q^2 \left(\sqrt{Q^2+r^2}-m\right)}{\kappa ^2 \left(Q^2+r^2\right)^{5/2}}.
\end{eqnarray}
In general, there is no relation between the fluid components. Considering that part of this fluid is electromagnetic and a second part is scalar, we verify that a third part of this fluid is given by
\begin{equation}
\rho^R=\frac{2 \gamma  Q^2 \left(\sqrt{Q^2+r^2}-3 m\right)}{\kappa ^2 \left(Q^2+r^2\right)^{5/2}}=-p_r^R=-p_t^R,
\end{equation}
which is an isotropic fluid with an equation of state $\rho=-p$.

Thus, if we want the Simpson-Visser solution to be derived from the Rastall gravity equations, we need to consider an anisotropic fluid, which can be decomposed into an electromagnetic part, a scalar part, and a third isotropic part.

\subsection{Bardeen-type}
Considering know the Bardeen-type spacetime, the components of the anisotropic fluid are written as
\begin{eqnarray}
    \rho=\frac{2 \gamma  Q^2 \left(\left(Q^2+r^2\right)^{3/2}-m \left(2 Q^2+r^2\right)\right)}{\kappa ^2 \left(Q^2+r^2\right)^{7/2}}+\frac{Q^2 \left(\frac{8 m
   r^2}{\left(Q^2+r^2\right)^{7/2}}-\frac{1}{\left(Q^2+r^2\right)^2}\right)}{\kappa ^2},\\
    p_r=-\frac{2 \gamma  Q^2 \left(\left(Q^2+r^2\right)^{3/2}-m \left(2 Q^2+r^2\right)\right)}{\kappa ^2 \left(Q^2+r^2\right)^{7/2}}-\frac{Q^2 \left(4 m r^2+\left(Q^2+r^2\right)^{3/2}\right)}{\kappa ^2
   \left(Q^2+r^2\right)^{7/2}},\\
    p_t=\frac{2 \gamma  Q^2 \left(\left(Q^2+r^2\right)^{3/2}-m \left(2 Q^2+r^2\right)\right)}{\kappa ^2 \left(Q^2+r^2\right)^{7/2}}+\frac{Q^2 \left(-2 m Q^2+5 m r^2+\left(Q^2+r^2\right)^{3/2}\right)}{\kappa ^2
   \left(Q^2+r^2\right)^{7/2}}.
\end{eqnarray}
Once again, we obtain that there is no relation between the fluid components. Considering that part of this fluid is electromagnetic and a second part is scalar, we verify that a third part of this fluid is given by
\begin{equation}
\rho^R=\frac{2 \gamma  Q^2 \left(\left(Q^2+r^2\right)^{3/2}-m \left(2 Q^2+r^2\right)\right)}{\kappa ^2 \left(Q^2+r^2\right)^{7/2}}=-p_r^R=-p_t^R,
\end{equation}
which is an isotropic fluid with an equation of state $\rho=-p$.

Thus, just like in the Simpson-Visser case, the Bardeen-type solution can be found in Rastall gravity if we consider a fluid with three components. Two of them anisotropic, scalar and electromagnetic, and one isotropic.

The quantities related to the phantom scalar field and nonlinear electrodynamics, both for the Simpson-Visser metric and the Bardeen-type solution, are the same as those obtained in the context of general relativity \cite{Rodrigues:2023vtm}.
\section{Conclusions}
In this comment, we show that the obtained electromagnetic Lagrangian contains an error in the term involving the Rastall gravity correction. We demonstrate that, even with the corrected Lagrangian, the electromagnetic functions exhibit inconsistencies. Despite
$L(F)$ and $L_F(F)$ being obtained independently, they must be related by the consistency relation \eqref{cons}, and we find that neither the results from \cite{Atazadeh:2023wdw} nor ours match. This occurs because when considering black bounce solutions, such as Simpson-Visser or Bardeen-type, in Rastall gravity, the phantom scalar field and nonlinear electrodynamics are not sufficient to describe the material content of these spacetimes.

Instead of starting from a combination of the scalar field with nonlinear electrodynamics, we map the material content of these solutions considering the interpretation of an anisotropic fluid. If we insist that part of this fluid is composed of a scalar fluid and an electromagnetic fluid, a third part is necessary in the composition of this fluid. This third contribution is actually an isotropic fluid.

Thus, we emphasize once again that this type of solution cannot be solely mapped as a simple combination of phantom scalar field and nonlinear electrodynamics when considering Rastall gravity.

\section*{Acknowledgments}
M.E.R. thanks Conselho Nacional de Desenvolvimento Cient\'ifico e Tecnol\'ogico - CNPq, Brazil  for partial financial support. M.S. would like to thank Funda\c c\~ao Cearense de Apoio ao Desenvolvimento Cient\'ifico e Tecnol\'ogico (FUNCAP) for partial financial support.


\end{document}